\documentclass[aps,prb,twocolumn,showpacs,groupedaddress]{revtex4}  %% REVTeX 4.0
\usepackage{graphicx}

\begin{document}

\title{Effect of local pumping on random laser modes}

\author{X. Wu, J. Andreasen, H. Cao}
%% for REVTeX4, each author name can be set in a separate \author{} field

\affiliation{Department of Physics and Astronomy, Northwestern University, Evanston, Illinois, 60208.}

\author{A. Yamilov}
\affiliation{Department of Physics, University of Missouri--Rolla, Rolla, MO 65409.}
%\email{h-cao@northwestern.edu}

\begin{abstract}
We have developed a numerical method based on the transfer matrix to calculate the quasimodes and lasing modes in one-dimensional random systems. Depending on the relative magnitude of the localization length versus the system size, there are two regimes in which the quasimodes are distinct in spatial profile and frequency distribution. In the presence of uniform gain, the lasing modes have one-to-one correspondence to the quasimodes in both regimes. Local excitation may enhance the weight of a mode within the gain region due to local amplification, especially in a weakly scattering system.
\end{abstract}

\pacs{42.55.Zz,42.25.Dd}

\maketitle 

\section{Introduction}
Random laser, in which optical feedback is provided by scattering of light due to spatial inhomogeneity of the medium rather than by well defined mirrors, has recently attracted much attention \cite{cao_WRM}. One important topic of research is the nature of random laser modes. For a random laser with non-resonant feedback, the lasing modes are the diffusive modes, i.e, the eigenmodes of the diffusion equation \cite{letokhov}. For a random laser with resonant feedback, the lasing modes are believed to be the quasimodes, i.e, the eigenmodes of the Maxwell equations \cite{cao_2000}. This belief implies the quasimodes of a passive random system are not modified by the presence of gain. Such assumption is confirmed by the numerical studies of lasing modes in the localization regime \cite{vanneste_2001,jiang_2002}. With the introduction of gain, the localized modes of a passive random system are preserved and serve as the lasing modes. This conclusion is extended to the random systems far from the localization regime without direct confirmation. The lasing modes are regarded as the quasimodes with small decay rate, in particular the anomalously localized states \cite{apalkov_2002,cao_2002}. However, recent theoretical study \cite{deych_2005} reveals that the quasimodes of a passive random system are not the genuine normal modes of the same system with gain. This is because the spatial inhomogeneity of dielectric constant introduces a linear coupling between the quasimodes, mediated by the polarization of the active medium. The latest development of semiclassical laser theory for open complex or random media leads to the speculation that the lasing mode in a weakly scattering system may be a composite of many quasimodes with low quality factor \cite{tureci_2006,tureci_2007}. Moreover, under local excitation the reabsorption outside the local gain region suppresses the feedback from the unpumped part of the random sample and effectively reduces the system size \cite{yamilov_2005}. The lasing modes are therefore completely different from the quasimodes and confined in the vicinity of the pumped region. All these studies prompt us to investigate carefully the relation between the lasing modes and the quasimodes in both global pumping and local pumping. In this paper, we address the question whether the lasing modes are the quasimodes of passive random systems. The answer to this question determines whether the statistical distribution of the decay rates of quasimodes can be used to predict the lasing threshold and the number of lasing modes for random laser \cite{misir_1998,ling_2001,patra_2003,chang_2003,apalkov_2005,hacken_2005,kottos_2005,lagendijk_2006}. 

We conduct detailed numerical studies of quasimodes and lasing modes in one-dimensional (1D) random systems. A numerical method based on the transfer matrix is developed to calculate the quasimodes as well as the lasing modes in the presence of global or local gain. The main advantage of this method as compared to the finite-difference time-domain method is that it can calculate the quasimodes of weakly scattering systems which overlap spectrally and have short lifetime. In our numerical simulation, the scattering strength is varied over a wide range. The quasimodes, as well as the lasing modes, are formed by distributed feedback in the random system. The conventional distributed feedback (DFB) laser, made of periodic structures, operates either in the over-coupling regime or the under-coupling regime \cite{kogelnik_1971}. The random laser, which can be considered as randomly distributed feedback laser, also has these two regimes of operation. In the under-coupling regime the system size $L$ is much less than the localization length $\xi$, while in the over-coupling regime $L>\xi$.  The dominant  mechanism for the mode formation differs in these two regimes, leading to distinct characteristics of mode profile and frequency distribution. With the introduction of uniform gain, the lasing modes have one-to-one correspondence to the quasimodes in both regimes. However, local pumping can make the lasing modes significantly different from the quasimodes, especially in the under-coupling systems. Some quasimodes even fail to lase no matter how high the pumping level is. The results we obtain help understanding the random lasing with resonant feedback in the weakly scattering systems \cite{frolov_1999}, especially the recent observations of periodic lasing peaks in frequency \cite{polson_2005,wu_2006}.

\section{Numerical method}

We have developed a numerical method based on the transfer matrix to compute the quasimodes of 1D passive systems. This time-independent method is also applied to the calculation of lasing modes at the threshold under global or local excitation. The random system is 1D layered structure. It is composed of $N$ dielectric layers with air gaps in between. The refractive index of the dielectric layers is $n_d$, and that of the air gaps is 1. Both the thickness $d_1$ of the dielectric layers and the thickness of air gaps $d_2$ are randomized. $d_{1,2}=\bar{d}_{1,2}(1+\sigma \eta)$, where $0<\sigma<1$ represents the degree of randomness, and $\eta$ is a random number in $\left[-1,1\right]$, $\bar{d}_{1}$  ($\bar{d}_{2}$) is the average thickness  of the dielectric layers (air gaps). Outside the random system the refractive index is constant and its value is equal to the average refractive index $n_{eff}$ of the random system to eliminate the boundary reflection. 

According to the transfer matrix formula:
\begin{equation}
\left(\begin{array}{c}p_{1}\\q_{1}\end{array}\right)= M \left(\begin{array}{c}p_{0}\\q_{0}\end{array}\right)
\label{transfer_matrix}
\end{equation}
where $p_0$ and $q_0$ represent the forward and backward propagating waves on one side of the random system, $p_1$ and $q_1$ on the other side, $M$ is a $2\times2$ transfer matrix that characterizes wave propagation through the random system. The eigenmode of such an open system can be defined as ``natural mode'' or ``quasimode'', which generalizes
the concept of eigenmode of a closed system \cite{dutra_2000}. It satisfies the boundary condition that there are no incoming waves but only outgoing waves through the boundary of a random system, namely $p_{0}=0$ and $q_{1}=0$. In a passive system (without gain or absorption, the refractive indices being real numbers), such boundary condition requires the vacuum wavevector be a complex number, $k_0 = k_{0r} + i k_{0i}$. Substituting the boundary condition into Eq. \ref{transfer_matrix}, we get $M_{22}=0$. Since $M_{22}$ is a complex number, both the real part and imaginary part of $M_{22}$ are equal to 0. These two equations are solved to find $k_{0r}$ and $k_{0i}$. $k_{0r}=\omega /c$ tells the frequency $\omega$ of a quasimode, and $k_{0i}= - \gamma /c $ gives the decay rate $\gamma$ of a quasimode. 

After finding $k_{0}$ of a quasimode, the corresponding wavefunction can be obtained by calculating the electric field distribution $E(x)$ throughout the random system with the transfer matrix $M(k_0)$. The wavefunction inside the random system can be written as $E(x)= E_{+}(x)e^{in(x)k_{0}x}+E_{-}(x)e^{-in(x)k_{0}x}$, where $n(x)$ is the (real part of) refractive index at position $x$, $E_{+}(x)e^{in(x)k_{0}x}$ represents the forward-propagating field, and $E_{-}(x)e^{-in(x)k_{0}x}$ the backward-propagating field. Since $k_{0}$ is a complex number, the amplitudes of forward and backward propagating fields are $E_{+}(x)e^{-n(x)k_{0i}x}$ and $E_{-}(x)e^{n(x)k_{0i}x}$ ($k_{0i}<0$). These expressions show that there are two factors determining the wavefunction. The first is $E_{\pm}(x)$, which originates from the interference of multiply-scattered waves. The second is $e^{\pm n(x) k_{0i}x}$, which leads to exponential growth of the wavefunction toward the system boundary. Outside the random system, the wavefunction grows exponentially to infinity due to the negative $k_{0i}$. This is clearly unphysical. Thus we disregard the wavefunction outside the random system and normalize the wavefunction within the random system to unity. 

Optical gain is introduced to the random system by adding an imaginary part $n_i$ (negative number) to the refractive index. In the case of uniform gain, $n_i$ is constant everywhere inside the system. Outside the random system $n_i$ is set to zero. Different from the quasimode of a passive system, the vacuum wavevector $k_0$ of a lasing mode is a real number. The wavevector inside the random system is a complex number, $k=k_r+i k_i= k_0 [n(x)+ i n_i]$. Its imaginary part $k_i= k_0 n_i$ is inversely proportional to the gain length $l_g$. The onset of lasing oscillation corresponds to the condition that there are only outgoing waves through the boundary of the random system. The absence of incoming waves requires $M_{22} = 0$ in Eq. (1). Again since $M_{22}$ is a complex number, both its real part and imaginary part are zero.  These two equations are solved to find $k_{0}$ and $n_{i}$. Each set of solution $(k_0, n_i)$ represents a lasing mode. $k_{0}=\omega /c$ sets the lasing frequency $\omega$, and $n_{i}k_{0}=k_i = 1/l_g$ gives the gain length $l_g$ at the lasing threshold. The spatial profile of the lasing mode is then obtained by calculating the field distribution throughout the random system with the transfer matrix $M(k_0, n_i)$. Since our method is based on the time-independent wave equation, it holds only up to the lasing threshold \cite{jiang_1999}. In the absence of gain saturation, the amplitude of a lasing mode would grow in time without bound. Thus we can only get the spatially-normalized profile of a lasing mode at the threshold. The lasing mode is normalized in the same way as the quasimode for comparison. The amplitudes of forward and backward propagating fields of a lasing mode are $E_{+}(x)e^{-n_i k_{0}x}$ and $E_{-}(x)e^{n_i k_{0}x}$ ($n_i < 0$). The exponential growth factors $e^{\pm n_{i} k_0 x}$ depend on the gain value $|n_{i} k_0|$.

Local pumping is commonly used in the random laser experiment. To simulate such situation, we introduce gain to a local region of the random system. Our method can be used to find the lasing modes with arbitrary spatial distribution of gain. The imaginary part of the refractive index $n_i(x) = \tilde{n}_{i} f(x)$, where $f(x)$ describes the spatial profile of gain and its maximum is set to 1, $\tilde{n}_{i}$ represents the gain magnitude. The lasing modes can be found in the way similar to the case of uniform gain. The solution to $M_{22} =0$ gives the lasing frequency $k_0$ and threshold gain $\tilde{n}_{i} k_0$. The normalized spatial profile of a lasing mode is then computed with $M(k_0, \tilde{n}_{i})$.  

\section{Results and discussion}

Using the method described in the previous section, we calculate the quasimodes of 1D random systems. The quasimodes are formed by distributed feedback from the randomly-positioned dielectric layers.  We investigate many random structures with different scattering strengths. Depending on the relative values of the localization length $\xi$ and the system length $L$, there are two distinct regimes in which the quasimodes are dramatically different: (i) over-coupling regime $L > \xi$; (ii) under-coupling regime $L \ll \xi$. 

As an example, we consider the random structure with $\bar{d}_1=100$nm and $\bar{d}_2=200$nm. $\sigma = 0.9$ for both $d_1$ and $d_2$. To change from the under-coupling regime to the over-coupling regime, we increase the refractive index $n_d$ of the dielectric layers. In particular, we take $n_d = 1.05$ and $2.0$. The larger $n_d$ leads to stronger scattering and shorter localization length $\xi$. To obtain the value of $\xi$, we calculate the transmission $T$ as a function of system length $L$. $\langle \ln T \rangle$ is obtained from averaging over 10,000 configurations with the same $L$ and $\sigma$. When $L > \xi$, $\langle \ln T(L) \rangle$ decays linearly with $L$, and $\xi^{-1} = -d \langle \ln T(L) \rangle / dL$.  In the wavelength ($\lambda$) range of 500nm to 750nm, $\xi$ exhibits slight variation with $\lambda$ due to the residual photonic bandgap effect. For $n_d= 1.05$, $\xi \sim 200-240 \mu$m, while for $n_d= 2.0$, $\xi \sim 1.2-1.5 \mu$m. In the calculation of quasimodes, we fix the number of dielectric layers $N = 81$ and $\langle L \rangle = 24.1 \mu$m. For $n = 1.05$, $\xi \gg L$ in the wavelength range of interest, thus the random system is in the under-coupling regime. In contrast, for $n = 2.0$, $ \xi \ll L$ and the system in the over-coupling regime.  

To illustrate the difference between over-coupling regime and under-coupling regime, we compare the quasimodes of the same random structure with different $n_d$, namely, $n_d= 2.0$ or $1.05$. Figures 1(a) and (b) are the typical transmission spectra of these two systems. For the system with $n_d=2.0$ most transmission peaks are narrow and well separated in frequency, while for $n_d=1.05$ the transmission peaks are typically broad and overlapped. We find $k_0 = k_{0r} + i k_{0i}$ of the quasimodes in the wavelength range of 500-750nm. Figure 1(c) shows the values of $k_{0r}$ and $k_{0i} / \langle k_{0i} \rangle$ of these modes ($\langle k_{0i} \rangle$ is the average over all the quasimodes in the wavelength range of 500-750nm). In the system with $n_d=2.0$, most quasimodes are well separated spectrally, and they match the transmission peaks. $k_{0r}$ corresponds to the frequency of a transmission peak, and $k_{0i}$ to the linewidth of a transmission peak. However, some quasimodes are located close to the system boundary, thus having relatively large $k_{0i}$. They are usually invisible in the transmission spectrum due to spectral overlap with neighboring transmission peaks, which cause the number of transmission peaks [Fig. 1(a)] slightly less than the number of quasimodes [solid squares in Fig. 1(c)]. In the system with $n_d=1.05$, however, the number of peaks or maxima in the transmission spectrum [Fig. 1(b)] is significantly less than the number of quasimodes [open circles in Fig. 1(c)]. This is because in the under-coupling regime the decay rates of the quasimodes often exceed the frequency spacing to neighboring modes. The spectral overlap of the quasimodes makes the transmission peaks less evident and some even buried by the neighboring ones. 
\begin{figure}
\centerline{\scalebox{1.0}{\includegraphics{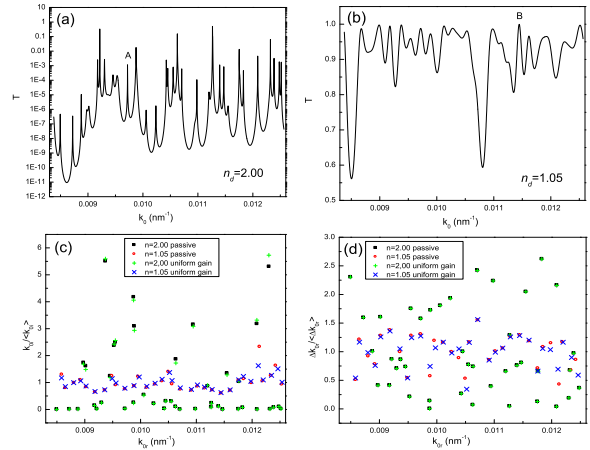}}}
\caption{(Color online) (a), (b): Transmission $T$ through a 1D random structure with $n_d=2.0$, $1.05$ as a function of vacuum wavevector $k_0$. (c) Frequencies $k_{0r}$  and normalized decay rates $k_{0i}/ \langle k_{0i} \rangle$ of the quasimodes in the random systems with $n_d=2.0$ (solid square) and $n_d=1.05$ (open circle), compared with the frequencies $k_0$ and normalized threshold gain $k_i/ \langle k_i \rangle$ of lasing modes in the same systems with $n_d=2.0$ ($+$) and $n_d=1.05$ ($\times$) under uniform excitation. (d) Normalized frequency spacing $\Delta k_{0r}/ \langle \Delta k_{0r} \rangle$ of neighboring quasimodes in the random systems with $n_d=2.0$ (solid square) and $n_d=1.05$ (open circle), compared with the normalized frequency spacing $\Delta k_{0}/ \langle \Delta k_{0} \rangle$ of neighboring lasing modes in the same systems with $n_d=2.0$ ($+$) and $n_d=1.05$ ($\times$) under uniform excitation.}
\label{fig1}
\end{figure}
It is clear in Fig. 1(c) that the decay rate fluctuation is much stronger in the random system with $n_d=2.0$ (solid squares) than that with $n_d=1.05$ (open circles). This is consistent with the broadening of quasimode decay rate distribution as a system approaches the localization regime with increasing scattering strength. Figure 1(d) plots the frequency spacing $\Delta k_{0r}$ between adjacent quasimodes normalized to the average value $\langle \Delta k_{0r} \rangle$.  The quasimodes of the random system with $n_d=1.05$ are more regularly spaced in frequency than those in the system with $n_d=2.0$. The average mode spacing is inversely proportional to the system length $L$.     

To interpret this phenomenon, we investigate the wavefunctions of the quasimodes. Figure 2(a) [(b)] shows the spatial distribution of intensity $I(x)= |E(x)|^2$ for a typical quasimode of the random system with $n_d=2.0$ ($n_d=1.05$). $I(x)$ is normalized such that the spatial integration of $I(x)$ within the random system is equal to unity.  The expression of $E(x)$ given in the previous section reveals the two factors determining the envelop of the wavefunction, i.e., the interference term $E^{\pm}(x)$ and the exponential growth term $e^{\pm n(x) k_{0i}x}$. Depending on which term is dominant, the spatial profile of the quasimodes can be drastically different. In the over-coupling regime, strong scattering makes the interference term dominant, and $I(x)$ exhibits strong spatial modulation. Most quasimodes are localized inside the random system, similar to the mode in Fig. 2(a). Their decay rates are low as a result of the interference-induced localization. In the under-coupling regime, the interference effect is weak due to small amount of scattering. The exponential growth term $e^{\pm n(x) k_{0i}x}$ dominates $E(x)$, making $I(x)$ increase exponentially towards the boundaries. The interference term only causes weak and irregular intensity modulation. A typical example of such mode profile is exhibited in Fig. 2(b). Since the quasimodes in the under-coupling system are spatially extended across the entire random system, the rates of light leakage through the boundaries are much higher than those of the localized modes in the over-coupling system.
\begin{figure}
\centerline{\scalebox{1.0}{\includegraphics{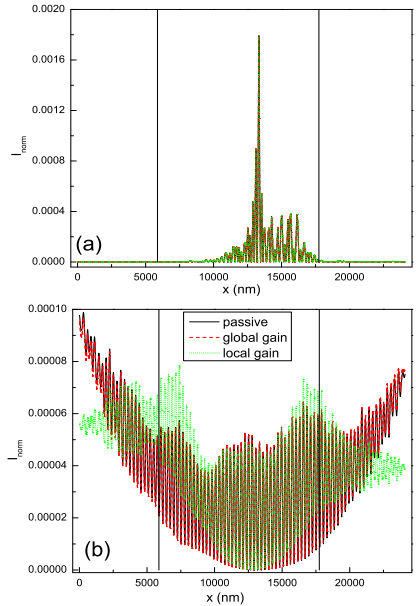}}}
\caption{(Color online) Spatial intensity distributions of quasimodes (black solid curve) and the corresponding lasing modes in the presence of global gain (red dashed curve) or local gain (green dotted curve). The pumped region is between the two vertical lines, $L_{p}=11.87\mu$m. (a) The mode marked as A in Fig. 1(a), $\lambda=646$nm, $n_d=2.0$. (b) The mode marked as B in Fig. 1(b), $\lambda=549$nm, $n_d=1.05$.}
\label{fig2}
\end{figure}
We repeat the above calculations with many random systems, and find the two different types of quasimodes are rather typical for the systems in the over-coupling and under-coupling regimes. The mode profiles and frequency spacings in the under-coupling systems reveal the feedbacks from the dielectric layers close to the boundaries are dominant over those from the interior. Thus the quasimodes in the under-coupling systems are formed mainly by the feedbacks from the scatterers near the system boundaries. However, the feedbacks from the scatterers in the interior of the system are weak but not negligible, e.g. they induce small fluctuations in the frequency spacings and the decay rates. Note that a random system in the under-coupling regime cannot be approximated as a uniform slab with the average refractive index $n_{eff}$, despite its quasimodes exhibit similar features as the Fabry-Perot modes formed by the reflections from the slab boundaries. Since in our calculation the refractive index outside the random system is set to $n_{eff}$, there would be no quasimodes if the random system were replaced by a dielectric slab of $n_{eff}$. Hence, the quasimodes in the under-coupling regime are not formed by the boundary reflection. In the over-coupling regime, the feedback from the scatterers deep inside the system becomes dominant, and the interference of multiply scattered waves lead to spatial localization of the quasimodes.   

Next we study the lasing modes in the random system with uniform gain and compare them to the quasimodes. $n_i$ is constant everywhere within the random system, so that the gain length $l_g = 1/ k_i= 1/k_0 n_i$ in the dielectric layers is equal to that in the air gaps. Using the method described in the previous section, we find the frequency and threshold gain of each lasing mode. We calculate the lasing modes in the same random systems as in Fig. 1 within the same wavelength range (500-750nm). The frequency $k_0$ and normalized threshold $k_i/ \langle k_i \rangle$ of each lasing mode are plotted in Fig. 1(c) for comparison with the quasimodes. It is clear that there exists one-to-one correspondence between the lasing modes and the quasimodes for the random systems in both over-coupling and under-coupling regimes. For the system with $n_d=2.0$, the lasing modes match well the quasimodes, with only slight difference between $k_i/ \langle k_i \rangle$ and $k_{0i}/ \langle k_{0i} \rangle$ for the relatively leaky modes. For the system with $n_d=1.05$, the deviation of the lasing modes from the quasimodes is more evident, especially for those modes with large decay rates. Such deviation can be explained by the modification of transfer matrix $M$. In the passive system, $k_{0i}$ is constant but $k_i = k_{0i} n(x)$ varies spatially. With the introduction of uniform gain, $k_i$ becomes constant within the random system, and the feedback inside the random system is caused only by the contrast in the real part of the wavevector $k_r = k_0 n(x)$ between the dielectric layers and the air gaps. With a decrease in the scattering strength, $k_{0i}$  in the passive system gets larger, and the ratio of the feedback caused by the contrast in $k_i$ to that in $k_r$ increases. The addition of uniform gain results in a bigger change of $M$, as it removes the feedback due to the inhomogeneity of $k_i$. Moreover, since there is no gain outside the random system, $k_i$ suddenly drops to zero at the system boundary. This discontinuity of $k_i$ generates additional feedback for the lasing modes. In the weakly scattering system, the threshold gain is high. The larger drop of $k_i$ at the system boundary makes the additional feedback stronger. To check its contribution to lasing, we replace the random system with a uniform slab of $n_{eff}$ while keeping the same gain profile. Since the real part of the refractive index or $k_r$ is homogeneous throughout the entire space, the feedback comes only from the discontinuity of $k_i$ at the slab boundaries. We find the lasing threshold in the uniform slab is significantly higher than that in the random system with $n_d=1.05$. This result confirms that for the random systems in Fig. 1, the additional feedback caused by the $k_i$ discontinuity at the system boundary is weaker than the feedback due to the inhomogeneity of $k_r$ inside the random system. However, if we further reduce $n_d$ or $L$, the threshold gain increases, and the feedback from the system boundary due to gain discontinuity eventually plays a dominant role in the formation of lasing modes.  

We also compute the intensity distribution $I(x)$ of each lasing mode at the threshold. $I(x)$ is normalized such that its integration across the random system is equal to 1. Such normalization facilitates the comparison of the lasing mode profile to the quasimode profile. In Fig. 2(a) [(b)], $I(x)$ of the lasing mode is plotted together with that of the corresponding quasimode. Although the lasing mode profiles in Figs. 2(a) and (b) are quite different, they are nearly identical to those of the quasimodes. For the localized mode in the random system with $n_d=2.0$, $I(x)$ of the lasing mode does not exhibit any visible difference from that of the quasimode in Fig. 2(a). For the extended mode in the system with $n_d=1.05$, the lasing mode profile deviates slightly from the quasimode profile, especially near the system boundaries. This deviation results from the modification of the transfer matrix $M$ by the introduction of uniform gain across the random system. The modification is bigger in the under-coupling system, leading to larger difference in the mode profile.  

Finally we investigate the lasing modes under local excitation. In particular, $f(x) = 1$ for $|x-x_c| \leq L_1/2$, $f(x)=\exp[-|x-x_c|/L_2]$ for $L_1/2 <|x-x_c| \leq L_1/2 +2L_2$, and $f(x)=0$ elsewhere. The lasing mode frequency $k_0$, the threshold gain $\tilde{k}_i = k_0 \tilde{n}_i$, and the spatial profile $I(x)$ are calculated with the method described in the previous section. $I(x)$ is normalized in the same way as that of quasimode for comparison. As an example, we consider the same random structures as in Fig. 1 and introduce gain to the central region $x_c=L/2$ of length $L_{p}= L_1+ 4L_2 =8.84 + 3.03 = 11.87\mu$m (marked by two vertical lines in Fig. 2). Figures 3(a) plots $k_0$ and $\tilde{k}_{i}/ \langle \tilde{k}_{i} \rangle$ for all the lasing modes within the wavelength range of 500-750nm. Comparing with Fig. 1, we find some quasimodes fail to lase under local pumping, no matter how high the pumping level is. The rest modes lase but their wavefunctions can be significantly modified by the particular local excitation. The two modes shown in Fig. 2 both lase under the local pumping configuration we consider. Their intensity distributions are plotted in Fig. 2. The mode in Fig. 2(a) is localized within the pumped region, and its spatial profile is barely modified by the local gain. In contrast, the mode in Fig. 2(b) is spatially extended and has less overlap with the central gain region. The intensity distribution of the lasing mode differs notably from that of the quasimode. The exponential growth of $I(x)$ towards the system boundaries is suppressed  outside the gain region, while inside the gain region $I(x)$ grows exponentially towards the ends of the gain region at a rate higher than that of the quasimode. These behaviors can be  explained by the spatial variation of gain. Outside the pumped region, there is no optical amplification thus light intensity does not increase exponentially. Within the pumped region, the faster intensity growth results from the higher threshold gain for lasing with local pumping than that with global pumping. Nevertheless, the close match in the number and spatial position of intensity maxima justifies the correspondence of the lasing mode to the quasimode.
\begin{figure}
\centerline{\scalebox{1.0}{\includegraphics{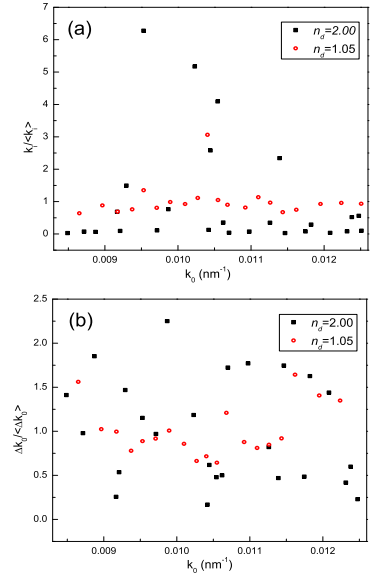}}}
\caption{(Color online) (a) Normalized threshold gain $\tilde{k}_{i}/\langle k_{i} \rangle$ versus the frequency $k_{0}$ of lasing modes in the random systems with $n_d=2.0$ (solid squares) and $n_d=1.05$ (open circles) under local excitation (between the two vertical lines in Fig. 2). (b) Normalized frequency spacing $\Delta k_{0}/\langle \Delta k_{0} \rangle$ of neighboring lasing modes in the systems with $n_d=2.0$ (solid squares) and $n_d=1.05$ (open circles) under local excitation.}
\label{fig3}
\end{figure}
We repeat the calculation with many modes under the same pumping configuration, and find the weight of a mode within the gain region is often enhanced. To quantify such enhancement, we introduce a parameter $\delta$ which is equal to the ratio of $I(x)$ integrated over the pumped region to that over the entire random system. We compare the values of $\delta$ for the lasing modes under local excitation to that of the corresponding quasimodes. For the mode Fig. 2(b), $\delta$ is increased from 0.33 for the quasimode to 0.41 for the lasing mode, while for the mode in Fig. 2(a) $\delta$ remains at 0.98. Thus the effect of local pumping is stronger for the modes in the weakly scattering system. This is because when scattering is weak the local gain required for lasing is high. The feedback within the pumped region is greatly enhanced, leading to the modification of mode profile. 

We also investigate the fluctuations in threshold gain and frequency spacing of lasing modes under local excitation. Figure 3(a) shows the lasing threshold fluctuation for the random system with $n_d=1.05$ is smaller than that with $n_d=2.0$. Since the number of lasing modes under local pumping is usually less than that of quasimodes, the average mode spacing $\langle \Delta k_0 \rangle$ is increased. Figure 3(b) plots the frequency spacing $\Delta k_0$ of adjacent lasing modes normalized to the average value $\langle \Delta k_0 \rangle$. There is more fluctuation in the mode spacing for the random system with $n_d=2.0$ than that with $n_d=1.05$. Hence, with local gain the frequency spacing of lasing modes is more regular in the under-coupling regime than in the over-coupling regime. This result is similar to that with uniform gain. 

Although the local pumping enhances the feedback within the pumped region, the feedback outside the pumped region cannot be neglected. To demonstrate this, we calculate the lasing modes in the reduced systems of length $L_{p}$ by replacing the random structures outside the gain region with a homogeneous medium of $n_{eff}$. The reduced system has uniform gain instead of the gain profile $f(x)$ in the original system. The results are shown in Fig. 4(a) for the system with $n_d=2.0$ and in Fig. 4(b) for the system with $n_d=1.05$.  The number of lasing modes in the reduced system is less than that in the original system under local pumping. In fact, the lasing modes are generally different, with only exception for a few modes localized within the gain region in the system with $n_d=2.0$. Moreover, the lasing threshold in the reduced system is higher than that in the original system with local gain. These differences are attributed to the feedbacks from the random structure outside the pumped region of the original system. It demonstrates the scatterers in the unpumped region also provides feedback for lasing. By comparing Figs. 4(a) and (b), we find the difference in the lasing threshold between the original system under local pumping and the reduced system is smaller for the system with $n_d=1.05$ than that with $n_d=2.0$. It indicates the contribution from the scatterers outside the gain region to lasing is reduced as the system moves further into the under-coupling regime. 

We note that local pumping introduces inhomogeneity in the imaginary part of the refractive index, which generates additional feedback for lasing. To check its effect, we simulate lasing in a homogeneous medium with the average refractive index $n_{eff}$. The local gain profile $f(x)$ remains the same. Only the spatial variation of $k_i(x)=k_0 \tilde{n}_i f(x)$ provides feedback for lasing. As shown in Figs. 4(a) and (b), the lasing thresholds are much higher than those in the random systems, even for the system with $n_d=1.05$. This result demonstrate that for the random systems in Figs. 3 and 4, the feedbacks for lasing under local pumping are predominately caused by the inhomogeneities in the real part of the refractive index $n(x)$ or the wavevector $k_r(x) = k_0 n(x)$. However, a further reduction in $n_d$ or $L_p$ could make the feedback due to the inhomogeneity of $k_i(x)$ significant.      
\begin{figure}
\centerline{\scalebox{1.0}{\includegraphics{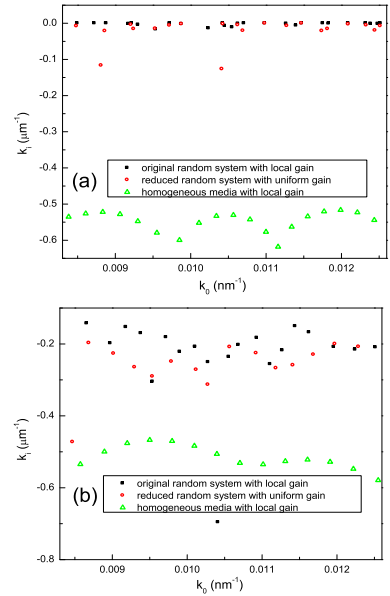}}}
\caption{(Color online): Threshold gain $\tilde{k}_{i}/ \langle \tilde{k}_{i} \rangle$ of lasing modes in the (original) random system of length 24.1$\mu$m with local excitation in the central region of length 11.87$\mu$m (solid square), compared to the threshold gain of lasing modes in the reduced system of length 11.87$\mu$m under uniform excitation (open circle) and the threshold gain of lasing modes in the homogeneous medium with $n_{eff}$ under local excitation in the region of length 11.87$\mu$m (a) $n_d=2.0$, $n_{eff}=1.3361$, (b) $n_d=1.05$, $n_{eff}=1.0168$.}
\label{fig4}
\end{figure}
\section{Conclusion}

We have developed a numerical method to calculate the quasimodes of 1D passive random systems and the lasing modes at the threshold with either global or local pumping. We identify two regimes for the quasimodes: over-coupling regime ($L > \xi$) and under-coupling regime ($L \ll \xi$). In the under-coupling regime the electric field of a quasimode grows exponentially towards the system boundaries, while in the over-coupling regime the field maxima are located inside the random system. The frequency spacing of adjacent modes are more regular in the under-coupling regime, and there is less fluctuation in the decay rate. The distinct characteristic of the quasimodes in the two regimes result from the different mechanisms of mode formation. In an over-coupling system, the quasimodes are formed mainly by the interference of multiply scattered waves by the particles in the interior of the random system. In contrast, the feedbacks from the scatterers close to the system boundaries play a dominant role in the formation of quasimodes in an under-coupling system. The contributions from the scatterers in the interior of the random system to the mode formation are weak but not negligible. They induce small fluctuations in mode spacing and decay rate. As the scattering strength is increased, the feedbacks from those scatterers in the interior of the system get stronger, and the frequency spacing of the quasimodes becomes more random. 

In the presence of uniform gain across the random system, the lasing modes (at the threshold) have one-to-one correspondence with the quasimodes in both over-coupling and under-coupling systems. However, the lasing modes may differ slightly from the corresponding quasimodes in frequency and spatial profile, especially in the under-coupling systems. This is because the introduction of uniform gain removes the feedback caused by spatial inhomogeneity of the imaginary part of the wavevector within the random system and creates additional feedback by the discontinuity of the imaginary part of the wavevector at the system boundaries. As long as the scattering is not too weak, the quasimodes are only slightly modified by the introduction of uniform gain to a random system and they serve as the lasing modes. This conclusion is consistent with that drawn from the time-dependent calculations \cite{vanneste_2001,jiang_2002,vanneste_2006}. Hence, the knowledge of the decay rates of the quasimodes, in conjunction with the gain spectrum, can predict the first lasing mode.  Because of the correspondence between the lasing modes and the quasimodes, the frequency spacing of adjacent lasing modes is more regular in the under-coupling systems with smaller mode-to-mode variations in the lasing threshold.   

When optical gain is introduced to a local region of the random system, some quasimodes cannot lase no matter how high the gain is. The rest modes can lase but their spatial profiles may be significantly modified. Such modifications originate from strong enhancement of feedbacks from the scatterers within the pumped region. It increases the weight of a lasing mode within the gain region. Nevertheless, the feedbacks from the scatterers outside the pumped region are not negligible. Moreover, the spatial variation in the imaginary part of the refractive index generates additional feedback for lasing. As the pumped region becomes smaller, the number of lasing modes is reduced, and the frequency spacing of lasing modes is increased. In an under-coupling system, the regularity in the lasing mode spacing remains under local excitation.
Our calculation results will help to interpret the latest experimental observations \cite{polson_2005,wu_2006} of spectral periodicity of lasing peaks in weakly scattered random systems under local pumping. We note that the effect of local excitation can be significant in an over-coupling system if the size of the pumped region is much smaller than the spatial extend of a localized mode or the spatial overlap between the pumped region and the localized mode is extremely small. Hence, caution must be exerted in using the decay rates of quasimodes to predict the lasing threshold or the number of lasing modes under local excitation. Finally we comment that the increase in the mode concentration in the gain region by local pumping have distinct physical mechanism from the absorption-induced localization of lasing modes in the pumped region \cite{yamilov_2005}. The former is based on selective enhancement of feedback within the gain region, while the latter on the suppression of the feedback outside the pumped region by reabsorption.

We acknowledge stimulating discussions with Profs. Christian Vanneste, Patrick Sebbah, and Lev I. Deych. This work is supported by the National Science Foundation under the Grant No. DMR-0093949.


\begin{thebibliography}{99}

\bibitem{cao_WRM} H. Cao, ''Lasing in random media,'' Waves in Random Media {\bf 13}, R1--R39 (2003). 

\bibitem{letokhov} V. S. Letokhov, ``Quantum statistics of multi-mode radiation from an ensemble of atoms,'' {\it Sov. Phys. JETP}  {\bf 26}, 1246--1251 (1968).

\bibitem{cao_2000} H. Cao H, J. Y. Xu, S. H. Chang, S. T. Ho, E. W. Seelig, X. Liu, and R. P. H. Chang, ''Spatial confinement of laser light in active random media,'' Phys. Rev. Lett. {\bf 84}, 5584--5587 (2000). 

\bibitem{vanneste_2001} C. Vanneste and P. Sebbah, ``Selective excitation of localized modes in active random media,'' Phys. Rev. Lett. {\bf 87}, 183903 (2001).

\bibitem{jiang_2002} X. Y. Jiang, and C. M. Soukoulis, ''Localized random lasing modes and a path for observing localization,'' Phys. Rev. E {\bf 65}, 025601 (2002).

\bibitem{apalkov_2002} V. M. Apalkov, M. E. Raikh, and B. Shapiro, ''Random resonators and prelocalized modes in disordered dielectric films,'' Phys. Rev. Lett. {\bf 89}, 016802 (2002).

\bibitem{cao_2002} H. Cao, Y. Ling, J. Y. Xu, and A. L. Burin, ''Probing localized states with spectrally resolved speckle techniques,'' Phys. Rev. E {\bf 66}, R025601 (2002).

\bibitem{deych_2005} L. I. Deych, ''Effects of spatial nonuniformity on laser dynamics,'' Phys. Rev. Lett. {\bf 95}, 043902 (2005).

\bibitem{tureci_2006} H. E. T\"{u}reci, A. D. Stone and B. Collier, ''Self-consistent multimode lasing theory for complex or random lasing media,'' Phys. Rev. A {\bf 74}, 043822 (2006).

\bibitem{tureci_2007} H. E. T\"{u}reci, A. D. Stone and L. Ge, ''Theory of the spatial structure of non-linear lasing modes''' arXiv:cond-mat/0610229.

\bibitem{yamilov_2005} A. Yamilov, X. Wu, H. Cao, and A. Burin, ''Absorption-induced confinement of lasing modes in diffusive random medium,'' Opt. Lett. {\bf 30}, 2430--2432 (2005).

\bibitem{misir_1998} T. Sh. Misirpashaev and C. W. J. Beenakker, ''Lasing threshold and mode competition in chaotic cavities,'' Phys. Rev. A {\bf 57} 2041--2045 (1998).

\bibitem{ling_2001} Y. Ling, H. Cao, A. L. Burin, M. A. Ratner, C. Liu, R. P. H. Chang, ''Investigation of random lasers with resonant feedback,'' Phys. Rev. A {\bf 64}, 063808 (2001).

\bibitem{patra_2003} M. Patra, ''Decay rate distributions of disordered slabs and application to random lasers,'' Phys. Rev. E, {\bf 67}, 016603 (2003).

\bibitem{chang_2003} S. H. Chang, H. Cao, and S. T. Ho, ''Cavity formation and light propagation in partially ordered and completely random one-dimensional systems,'' IEEE J. Quantum Electron. {\bf 39}, 364--374 (2003).

\bibitem{apalkov_2005} V. M. Apalkov and M. E. Raikh, ''Universal fluctuations of the random lasing threshold in a sample of a finite area,'' Phys. Rev. B, {\bf 71} 054203 (2005).

\bibitem{hacken_2005} G. Hackenbroich, ''Statistical theory of multimode random lasers,'' J. Phys. A: Math. Gen. {\bf 38}, 10537--10543 (2005).

\bibitem{kottos_2005} T. Kottos, ''Statistics of resonances and delay times in random media: beyond random matrix theory,'' 
J. Phys. A: Math. Gen. {\bf 38}, 10761--10786 (2005).

\bibitem{lagendijk_2006} K. L. van der Molen, A. P. Mosk, and A. Lagendijk, ''Intrinsic intensity fluctuations in random lasers'', Phys. Rev. A {\bf 74}, 053808 (2006).

\bibitem{kogelnik_1971} H. Kogelnik, and C. V. Shank, ''Coupled-Wave Theory of Distributed Feedback Lasers,'' J. Appl. Phys. {\bf 43}, 2327--2335 (1972).

\bibitem{frolov_1999} S. V. Frolov, Z. V. Vardeny, K. Yoshino, A. A. Zakhidov, and R. H. Baughman, ''Stimulated emission in high-gain organic media,'' Phys. Rev. B {\bf 59}, R5284--R5287 (1999).

\bibitem{polson_2005} R. C. Polson, and Z. V. Vardeny, ''Organic random lasers in the weak-scattering regime,'' Phys. Rev. B {\bf 71}, 045205 (2005).

\bibitem{wu_2006} X. Wu, A.  Yamilov, A. A. Chabanov, A. A. Asatryan, L. C. Botten, and H. Cao, ''Random lasing in weakly scattering systems,'' Phys. Rev. A, {\bf 74}, 053812 (2006).

\bibitem{dutra_2000} S. M. Dutra, and G. Nienhuis, ''Quantized mode of a leaky cavity,'' Phys. Rev. A {\bf 62}, 063805 (2000).

\bibitem{jiang_1999} X. Y. Jiang, and C. M. Soukoulis, ''Symmetry between absorption and amplification in disordered media,'' Phys. Rev. B {\bf 59}, R9007--R9010 (1999).

\bibitem{vanneste_2006} C. Vanneste, P. Sebbah, and H. Cao, unpublished. 

\end{thebibliography}
\end{document}